\documentstyle[12pt]{article}
\def\beq{\begin{equation}}
\def\eeq{\end{equation}}
\def\bea{\begin{eqnarray}}
\def\eea{\end{eqnarray}}
\def\bq{\begin{quote}}
\def\eq{\end{quote}}

\def\LNC{{\it Lett. Nuovo Cimento} }

\def\NC{{\it Nuovo Cimento} }

\def\PL{{\it Phys.Lett.} }
\def\PR{{\it Phys.Rev.} }
\def\PRL{{\it Phys.Rev.Lett.} }

\def\gappeq{\mathrel{\rlap {\raise.5ex\hbox{$>$}}
{\lower.5ex\hbox{$\sim$}}}}

\def\lappeq{\mathrel{\rlap{\raise.5ex\hbox{$<$}}
{\lower.5ex\hbox{$\sim$}}}}

\begin{document}
\pagestyle{empty}
\begin{flushright}
{CERN-TH/97-23}\\
\end{flushright}
\vspace*{5mm}
\begin{center}
{\bf A THEOREM ON THE REAL PART}\\
{\bf  OF THE HIGH-ENERGY SCATTERING AMPLITUDE} \\
{\bf  NEAR THE FORWARD DIRECTION} \\
\vspace*{1cm} 
{\bf Andr\'e MARTIN}$^{*)}$ \\
\vspace{0.3cm}
Theoretical Physics Division, CERN \\
CH - 1211 Geneva 23 \\
and \\ LAPP$^{*)}$, B.P. 110 \\ F - 74941 Annecy Le Vieux\\
{\tt E-mail:
martina@mail.cern.ch}\\

\vspace*{2cm}  
{\bf ABSTRACT} \\ \end{center}
\vspace*{5mm}
\noindent
We show that if for fixed negative (physical) square of the momentum transfer
$t$, the differential cross-section ${d\sigma\over dt}$ tends to zero and if
the total cross-section tends to infinity, when the energy goes to infinity,
the real part of the even signature amplitude cannot have a constant sign
near $t = 0$.
\vspace*{3cm}

\vspace*{0.1cm}
\noindent
\rule[.1in]{17cm}{.002in}
\noindent
$^{*)}$ URA 14-36 du CNRS, associ\'ee \`a L'Ecole Normale 
Sup\'erieure de Lyon et \`a l'Universit\'e de Savoie. 
\vspace*{0.1cm}

\begin{flushleft} CERN-TH/97-23 \\
January 1997 
\end{flushleft}
\vfill\eject

\setcounter{page}{1}
\pagestyle{plain}

The measurement of the real part of the scattering amplitude in the forward
direction is a crucial test of dispersion relations. It allows to get
information on the total cross-section at higher energies
\cite{aaa},\cite{bb} and also it may be a crucial test of locality. A
violation of dispersion relations might be the sign of new physics like for
example the existence of extra compact dimensions \cite{cc}.

To measure this real part, one has to use Coulomb interference, which is
large at a very small but not zero momentum transfer. Most of the time people
assume that in the region where the interference takes place, the ratio of
the real to imaginary part, called $\rho$, is constant. For exceptions, see the
papers of Kundrat and  collaborators \cite{dd}. The present paper contains a
theorem which constitutes a kind of warning against the dangers of this
procedure, because we show that under assumptions which are satisfied by most
existing models and seem to be compatible with experimental data on proton
(anti)proton scattering, namely that ${d\sigma\over dt}$ tends to zero for
fixed negative $t$ and that $\sigma_{\rm total}$ tends to infinity, the real
part of the even signature amplitude cannot have a constant sign in a strip
$-T < t \leq 0$, $s > s_M$, where $T$ is arbitrarily small and $s_M$ is
arbitrarily large. Hence, $\rho$ cannot  be constant because it is extremely
difficult to make the imaginary part of the amplitude vanish exactly at the
same place as the real part because of positivity constraints, and also
because, from a practical point of view, the differential cross-section is
not seen to vanish anywhere. The question is therefore if it is still
admissible to neglect the variation of
$\rho$ in a measurement of the real part by interference. In a simple case we
shall show that the zero of the real part is more than 100 times further away
than the place where the interference is maximum, but we have no certainty
about this in general. Of course, in a specific model, like for instance the
one of Wu and collaborators \cite{ee}, there is no problem because one can test
directly if
$\rho (s,t)$ given by the model reproduces the interference curve, but it is
unfortunate that one cannot get $ \rho(s, t=0)$ without any
theoretical prejudice.

We now come to Theorem I.
Consider an amplitude $F(s,t,u)$ describing $A+B \rightarrow A+B$ and $A+\bar
B \rightarrow A + \bar B$ reactions with
$$
s + t + u = 2M_A^2 + 2M^2_B~,
$$
$s$, square of centre-of-mass energy in the $A+B\rightarrow A+B$ channel and
$u$ in the $A+\bar B \rightarrow A + \bar B$ channel and $t = -2k^2(1-\cos
\theta )$, $k$ centre-of-mass momentum and $\cos\theta$ scattering angle in
the AB channel. If the amplitude is not symmetric in $s-u$ exchange, we
symmetrize it to get the ``even-signature" amplitude. Then we state:

\noindent
$\underline{\rm Theorem~~ I}$ \\
If $F$ is $s-u$ crossing symmetric and if Re$F$ has a constant sign (including
zero) in the strip $s_M~<~s~<~\infty$, $-t < T \leq 0$ where $s_M$ is arbitrarily
large and $T > 0$ arbitrarily small, 
and if ${F(s,t)\over s}\rightarrow 0$ for $-T < t < 0$,
$\vert~F(s,t~=~0)/s\vert$ does not tend to
$\infty$ for $s \rightarrow \infty$ and therefore by the optical theorem the
total cross-section does not tend to infinity.

\noindent
$\underline{\rm Theorem~~ II}$ \\
If ${d\sigma\over dt}~(A+B\rightarrow A+B)$ and 
${d\sigma\over dt}~(A+\bar B\rightarrow A+\bar B)$
tend to zero for $s\rightarrow\infty$ $-T<t < 0$ for some $T$ and if
$\sigma_{\rm total}(AB)$ and/or  $\sigma_{\rm total}(A\bar B)$
tend to infinity for infinite energy, the real part of the even signature
amplitude cannot have a constant sign in a strip $s_M < s < \infty$, $-T < t
\leq 0$, for any $s_M$ and $T > 0$.

Theorem II is an obvious consequence of Theorem I.

\noindent
$\underline{\rm Proof~ of~ Theorem~~ I}$ \\
We work with a crossing symmetric amplitude in the exchange $s
\leftrightarrow u$. For $-T < t \leq 0$ this amplitude satisfies twice
subtracted dispersion relations (in fact this holds also for $-T~<~t~<~+R$,
$R > 0$ as a consequence of the axioms of local field theory and positivity
\cite{ff}). It means that the amplitude is analytic in a twice cut plane. If
we use the variable
\beq
z(s,t) = (s-u)^2 = \left(2s - 2(M^2_A+M^2_B) + t\right)^2~,
\label{one}
\eeq
\beq
F(s,t,u) = G(z,t)~,
\label{two}
\eeq 
$G$ is analytic in a once cut plane for $-T < t$ real $\leq 0$, with, in the
simplest case of a ``normal threshold"
a cut beginning at
\beq
z_0 (t) = (4M_AM_B + t)^2
\label{three}
\eeq

We can write an ``inverse" dispersion relation for $G$. Define:
\beq
H = {G(z,t)\over \sqrt{z_0(t)-z}}~.
\label{four}
\eeq
If $F(s,t)/s$ tends to zero for $t < 0$, $s\rightarrow\infty$, $H$
satisfies an $\underline{\rm unsubtracted}$ dispersion relation
\beq
H(z,t) = -~{1\over\pi}~~ \int^\infty_{z_0(t)}~~{{\rm Re} G(z^\prime,t)\over
\sqrt{z^\prime -z_0}~(z^\prime -z)}~dz~.
\label{five}
\eeq
Suppose now that Re$F(s,t)$ has a constant sign, say negative, for $s > s_0$,
$-T < t \leq 0$. Re$G(z,t)$ has then also a constant negative sign for $-T <
t \leq 0$ and $z > z_M(t) = 
\left[ 2(s_M-M^2_A - m^2_B) + t\right]^2$

We write $H$ as
\beq
H = \hat H + \Delta
\label{six}
\eeq
where
\beq
\hat H_{(z,t)} = {1\over \pi}~~\int^\infty_{z_M(t)}~~{{-\rm Re}G(z^\prime
,t)dz^\prime \over \sqrt{z^\prime - z_0(t)}~~(z^\prime -z)}~,
\label{seven}
\eeq
and
\beq
\Delta (z,t) = {1\over\pi}~~\int^{z_M(t)}_{z_0(t)}~~{{-\rm Re}G(z^\prime
,t)dz^\prime \over \sqrt{z^\prime - z_0(t)}~~(z^\prime -z)}~.
\label{eight}
\eeq
Notice that for $\vert z\vert \rightarrow\infty$, $\vert\Delta\vert$ is bounded
by const$/\vert z\vert$. 

$\hat H$ having a positive discontinuity is a ``Herglotz" function, having
no zeros in the complex plane. For $z < 0$, it is positive and increasing.
Hence
\beq
0 < \hat H(-x,t) < \hat H(0,t)~~{\rm for}~~ -T < t < 0~,~~~~x > 0~,
\label{nine}
\eeq
so  that $\hat H$ is uniformly bounded on the negative real axis.

Now we remark that $F(s,t)$ is analytic in $t$, for fixed $s$ in a circle
$\vert t\vert \leq R$ \cite{ff}, and hence, $G(z,t)$ and $H(z,t)$ are both
analytic in a neighbourhood of $-R < t < +R$ for any $\underline{\rm
complex}$ $z$ and also for $z \leq 0$ (one must pay attention to the fact
that the cuts move with $t$).

Concerning $\Delta (z,t)$, we can reexpress it as an integral over the
original variables $s^\prime$ and $t$ over the range $s_0 < s^\prime
< s_M$, and reach again the conclusion that when $z$ is strictly outside the
cut
$$
\left(2(s_0-M^2_A-M^2_B)-T\right)^2 < z < \left(2(s_M-M^2_A-M^2_B)\right)^2~,
$$
$\Delta$ is analytic in $t$ in a neighbourhood of $-R < t < +R$.

Therefore, $\hat H$ is also analytic in $t$, for fixed $z$, $-\infty < z < 0$,
in a neighbourhood of $-R~<~t~<~+R$. Hence inequality (\ref{nine}) which was
established for $t$ strictly negative also holds by continuity for $t = 0$:
\beq
0 \leq \hat H (-x,0) \leq \hat H(0,0)
\label{ten}
\eeq

Now suppose that $\sigma_{\rm total}\rightarrow\infty$ for
$s\rightarrow\infty$. This means that $\vert
F(s,t=0)/s\vert\rightarrow\infty$, and hence 
$\vert
H(z,t=0)\vert\rightarrow\infty$ for $z \rightarrow +\infty$ on both sides of
the cut.

Since $\vert\Delta (z,t=0)\vert$ can be bounded by const$/\vert z\vert$, this
also means that
\beq
\vert\hat H(z,t=0)\vert \rightarrow \infty~.
\label{eleven}
\eeq

But $-1/\hat H(z,t=0)$ is also a Herglotz function and, from (\ref{eleven})
\beq
\bigg\vert {1\over\hat H(z,t=0)}\bigg\vert \rightarrow 0~,
\label{twelve}
\eeq
for $z \rightarrow +\infty$, on both sides of the cut. By the
Phragm\'en-Lindel\"of theorem $1/\hat H(z,t=0)$ also tends to zero for $\vert
z\vert\rightarrow\infty$ in any complex direction. This contradicts the fact
that $\hat H(-x,t=0)$ is uniformly bounded for $x > 0$, and therefore
$\sigma_t(s)$ cannot tend to infinity.

Strictly speaking this implies that $\sigma_t(s)$ has a finite least lower
limit for $s\rightarrow\infty$. One can probably make a more refined statement
on a smoothed $\sigma_t$. Of course, if we believe that $\sigma_t$ is
monotonous beyond a certain energy, it means that $\sigma_t$ is
$\underline{\rm bounded}$.

As we said, Theorem II is a direct consequence of Theorem I. In practice,
Theorem II is the most relevant, because there are clear indications, in the
case of proton-antiproton scattering that $\sigma_t$ is increasing \cite{ggg}
and, comparing for instance the ISR and ${\rm sp\bar ps}$ data \cite{ggg}, that
${d\sigma \over dt}$
decreases at least for $t < -0.4$ GeV$^2$, and it seems that the decrease
occurs for smaller and smaller $\vert t\vert$ when the energy increases.

The simplest way to satisfy theorem II is to have a curve in the $s-t$ plane
where the real part changes sign:
\beq
t = -f(s)~,~~~f(s) > 0
\label{thirteen}
\eeq
such that $f(s) \rightarrow 0$ for $s\rightarrow\infty$, but this is not the
only possibility. We could also have an infinite sequence of bubbles in the
$s-t$ plane approaching $t=0$ for $s\rightarrow\infty$ in which Re$F$ would
have an opposite sign. The implication of the theorem for the individual
$AB\rightarrow AB$ and $A\bar B \rightarrow A\bar B$ amplitudes is that their
real parts cannot have both the $\underline{\rm same}$ constant sign.

Now, is the theorem a surprise? Not really if we look at simple-minded
examples. What is a surprise is its generality.

Examples can be built by taking
\beq
\vert F(s,t)\vert \sim s^{1+\lambda t}(\log s)^\gamma
\label{fourteen}
\eeq
with $0 < \gamma \leq 1$, $\lambda > 0$.   $F/s$  goes to zero for $t <
0~~s\rightarrow\infty$, and $\sigma_t \sim (\log s)^\gamma$.

Such examples do not manifestly violate $s$ channel unitarity nor polynomial
boundedness for fixed complex $t$ (for $\sigma_t\sim (\log s)^\gamma$, $1 <
\gamma < 2$ model building is more tricky while the extreme case $\gamma = 2$
is easy to handle \cite{hh}).  By standard techniques we can make this
amplitude even under crossing by replacing it by
\beq
F(s,t) = i ~C~e^{{-i\lambda t\pi\over 2}}~s^{1+\lambda t}~~\left(\log
s-{i\pi\over 2}\right)^\gamma~~,
\label{fifteen}
\eeq
with $C$ real $> 0$.

For small $t$, we see that the real part changes sign for
\beq
 t = -{\gamma\over \lambda\log s}
\label{sixteen}
\eeq

The special case $\gamma =1$ corresponds to what is called ``geometrical
scaling". Indeed, $\phi~=~F(s,t)/F(s,0)$ is just a function of $t \log s$,
i.e., of $\tau = t \sigma_t$, since $\sigma_t\sim\log s$. In that special
case the real part, according to an old theorem of the author \cite{jj}, is
found proportional to ${d\over d\tau}\left(\tau\phi(\tau )\right)$ and we
check that this gives us precisely a zero at
\beq
t = -{1\over \lambda \log s}
\label{seventeen}
\eeq
However, for $\gamma < 1$ this is no longer true and this shows how dangerous
is the abusive use of this ``magic" formula.

We would like now to return to the question of how
our result affects the measurement of the real part. In general, it is
difficult to say, but, in the special case of geometrical scaling the same
function determines the point of maximal interference and the location of the
zero of the real part. Indeed the complete amplitude, including the Coulomb
term is given (with some oversimplifications !) by \cite{hh}
\beq
F = {\alpha s\over 2t} + F_H(s,t)~,
\label{eighteen}
\eeq
where $F_H$ reduces to $s \sigma_{\rm total}(\rho + i) / 16\pi$ at $t = 0$.
We can rewrite $F$ as
$$
{\alpha s\over 2t} + {s \sigma_t (\rho(0) +i)\over 16\pi}~~\phi(t \sigma_t)~,
$$
and  the relative maximum interference occurs near
\beq
\psi(t \sigma_t) = 8\pi \alpha~,
\label{nineteen}
\eeq
where
\beq
\psi(\tau ) = \tau \phi (\tau )~,
\label{twenty}
\eeq
while the zero of the real part is given by
$$
\psi^\prime = 0~,
$$

It happens that in practice the point of maximum relative interference
for proton-(anti)proton scattering corresponds to  $t~\simeq~10^{-3}$~GeV$^2$
while, if we approximate the diffraction peak by exp~$Bt$, the zero of
$\psi^\prime$ is at
$t = (B/2)^{-1}
\cong$ 0.13 (GeV)$^2$ at $\sqrt{s}$ = 1 TeV
\cite{ggg},\cite{dd}, so that the situation is not too bad. However, outside
the geometrical scaling regime (which, incidentally does not seem to apply to
proton-antiproton scattering at energies above the ISR range, since the ratio
of the elastic to total cross-sections increases instead of being constant
\cite{ggg}), Eq. (\ref{sixteen}) shows that the zero can be at a different
place. In the model of Wu et al. \cite{ee} the first zero of Re $F$ is at $t
= -0.32$ GeV$^2$ for $\sqrt{s}$ = 1 TeV, according to C. Bourrely.

Finally, for completeness we would like to present the counterpart of Theorem
I for the odd signature amplitude:

\noindent
$\underline{Theorem~III}$\\
If, in the strip, $-T < t < 0$, $s > s_M$, the difference of the imaginary
parts of the amplitudes $AB \rightarrow AB$ and $A\bar B \rightarrow A\bar B$
has a constant sign, and if ${d\sigma \over dt} \rightarrow 0$ for both
reactions, the difference of the total cross-sections $\sigma_{AB} -
\sigma_{A\bar B}$ does not tend to infinity.

This means that the ``maximal odderon" \cite{kk} is excluded under the above
conditions. We leave the proof as an exercise to the reader.

\noindent
$\underline {\bf Acknowledgements}$

I would like to thank Jan Fischer, Maurice Haguenauer and Claude Bourrely for
very stimulating discussions.

\vfill\eject

\end{document}